\documentclass[prl,twocolumn,superscriptaddress,floatfix,longbibliography]{revtex4-2}

\usepackage{amsmath,amssymb} 
\usepackage{bm} 
\usepackage{graphicx} 
\usepackage{comment} 
\usepackage{subfigure} 
\usepackage{physics}%

\usepackage{enumitem}
\setlist{noitemsep,leftmargin=*,topsep=0pt,parsep=0pt}

\usepackage{xcolor} 
\definecolor{lightgray}{gray}{0.6}
\definecolor{medgray}{gray}{0.4}

\usepackage{hyperref}
\hypersetup{
colorlinks=true,
urlcolor= blue,
citecolor=blue,
linkcolor= blue,
}

\newif\ifptitle
\newif\ifpnumber
\newcounter{para}

\ptitletrue  
\pnumbertrue  

\newcommand{\mytitle}{An entanglement-aware quantum computer simulation algorithm}

\begin{document}

\title{\mytitle}

\author{Maxime Oliva}
\email[]{maxime.oliva@eviden.com}
\affiliation{Eviden Quantum Laboratory, Les Clayes-sous-Bois, France}

\date{\today}

\begin{abstract}
The advent of quantum computers promises exponential speed ups in the execution of various computational tasks. 
While their capabilities are hindered by quantum decoherence, they can be exactly simulated on classical hardware at the cost of an exponential scaling in terms of number of qubits.
To circumvent this, quantum states can be represented as matrix product states~\cite{vidal_efficient_2003, schollwoeck_density-matrix_2011} (MPS), a product of tensors separated by so-called bond dimensions. Limiting bond dimensions growth approximates the state, but also limits its ability to represent entanglement. Methods based on this representation have been the most popular tool at simulating large quantum systems. But how to trust resulting approximate quantum states for such intractable systems sizes ? I propose here a method for inferring the fidelity of an approximate quantum state without direct comparison to its exact counterpart, and use it to design an ``entanglement-aware'' (EA) algorithm for both pure and mixed states. As opposed to state of the art methods which limit bond dimensions up to an arbitrary maximum value, this algorithm receives as input a fidelity, and adapts dynamically its bond dimensions to both local entanglement and noise such that the final quantum state fidelity at least reaches the input fidelity. I show that this algorithm far surpasses standard fixed bond dimension truncation schemes. In particular, a noiseless random circuit of 300 qubits and depth 75 simulated using MPS methods takes one week of computation time, while EA-MPS only needs 2 hours to reach similar quantum state fidelity.
\end{abstract}

\maketitle
\section{Introduction}

Tensor network methods have been among the most popular avenue at circumventing the exponential scaling of exact quantum simulations \cite{vidal_efficient_2003, schollwoeck_density-matrix_2011, white_quantum_2018, noh_efficient_2020, cheng_simulating_2021}. While lowly entangled pure states can be efficiently simulated as matrix product states\cite{vidal_efficient_2003, schollwoeck_density-matrix_2011}, mixed states can be simulated as matrix product operators\cite{schollwoeck_density-matrix_2011, noh_efficient_2020}. Both approaches have allowed the reach of systems sizes far beyond what exact computation can achieve.

Any pure state can be represented as a matrix product state (MPS). The state vector of a quantum state of $N$ qubits is cast into a factorized form of $N$ tensors connected to each other with what are generally called ``bond dimensions''. Physically, bond dimensions can be thought of as the amount of entanglement a quantum state can encapsulate\cite{vidal_efficient_2003}. Contracting every tensor along its bond dimensions gives back the quantum state in its standard state vector form. The usual approach to approximating MPS consists in limiting these bond dimensions up to a maximum value set arbitrarily prior to the simulation. The more entanglement, the bigger the truncation errors, which limits MPS effectiveness to lowly entangled, but possibly very large quantum states. Such methods suffer two limitations. First, there is no way to assess how well the resulting quantum state approximates the exact quantum state when the system size is out of exact simulation reach. Secondly, there is no general method for guessing which maximum bond dimension is best for a given problem.

I show that for both pure and mixed states simulations, an approximate quantum state fidelity with respect to its exact counterpart can be indirectly computed, that is without ever having to compute the exact quantum state itself. This computation is done in real-time throughout the simulation, by efficiently computing the fidelity of every bond dimension truncation.

Based on this result, an entanglement-aware simulation algorithm can be designed, fully leveraging the presence or absence of both noise and entanglement. Rather than limiting bond dimensions uniformly as is the case in standard algorithms, bond dimensions are allowed to increase or decrease solely based on how this affects the resulting quantum state fidelity. As such the algorithm does not receive as input a maximum bond dimension, but instead a fidelity the final quantum state has to at least reach. Decreasing (resp. increasing) the desired fidelity decides how aggressive (resp. conservative) the algorithm will be at truncating bond dimensions.

Since bond dimensions are kept as low as possible for every operation, the algorithm efficiency far surpasses that of regular fixed maximum bond dimensions algorithms, while simultaneously allowing trustworthy output quantum states.

\section{The matrix product formalism}\label{sec:matrix-product-formalism}
Any pure quantum state can be written as a state vector $\ket{\psi}$ defined by :
\begin{eqnarray}
	\ket{\psi} = \sum_{\sigma}C_{\sigma_1,...,\sigma_N}\ket{\sigma_1}\otimes...\otimes\ket{\sigma_N} \label{eq:qstate_def}
\end{eqnarray}
with $C_{\sigma_1,...,\sigma_N}$ a 1D tensor containing $2^N$ complex values and $\{\ket{\sigma_i}\}$ forming an orthonormal basis.

As described in Vidal's article\cite{vidal_efficient_2003}, $\ket{\psi}$ can be decomposed as an MPS via successive singular value decomposition (SVD) of $\ket{\psi}$ in Eq.~(\ref{eq:qstate_def}):
\begin{equation}
	\allowbreak
	\ket{\psi} = \sum_{\substack{\sigma,\chi}} A^{[1]\sigma_1}_{1,\chi_{1}}A^{[2]\sigma_2}_{\chi_{1},\chi_{2}}...A^{[N]\sigma_N}_{\chi_{N-1},1} \  \ket{\sigma_1}\otimes...\otimes\ket{\sigma_N} \label{eq:mps_repr}
\end{equation}

We obtain a product of $N$ complex-valued tensors $\{A^{[i]\sigma_i}\}$, separated by bond dimensions $\{\chi_1, \chi_2, ..., \chi_{N-1}\}$. 
Assuming the $\chi$ are bounded by a maximum bond dimension $\chi_{max}$, the number of values contained in the MPS in Eq.~(\ref{eq:mps_repr}) scales in $\mathcal{O}(N\chi_{max}^2)$ with $N$ the number of qubits of the system.

Similarly, for mixed state, instead of considering the state vector $\ket{\psi}$, we consider the density matrix $\hat{\rho}$:
\begin{equation}
	\allowbreak
	\hat{\rho} = \sum_{\substack{\sigma, \sigma'}}C_{\sigma_{1}\sigma'_{1},...,\sigma_{N}\sigma'_{N}}|{\sigma_1}\rangle\langle{\sigma'_1}|\otimes...\otimes|{\sigma_N}\rangle\langle{\sigma'_N}|
\end{equation}
with $C_{\sigma_{1}\sigma'_{1},...,\sigma_{N}\sigma'_{N}}$ a 2D tensor containing $2^N \times 2^N$ complex values and $\{\ket{\sigma_i}\}$ forming an orthonormal basis.
Again, $\hat{\rho}$ can be cast into a matrix product operator (MPO) \cite{schollwoeck_density-matrix_2011} via successive SVD decompositions:
\begin{align}
	\allowbreak
	\hat{\rho} = \sum_{\chi_1...\chi_N} A^{[1]\sigma_1, \sigma'_1}_{1, \chi_1}&A^{[2] \sigma_2, \sigma'_2}_{\chi_1, \chi_{2}} ...A^{[N] \sigma_N, \sigma'_N}_{\chi_{N-1}, 1}\nonumber\\
	& |{\sigma_1}\rangle\langle{\sigma'_1}|\otimes...\otimes|{\sigma_N}\rangle\langle{\sigma'_N}|
\end{align}
Limiting the growth of bond dimensions \mbox{$\chi \leq \chi_{max}$} approximates the state, and let the MPO scale in $\mathcal{O}(N^2\chi_{max}^3)$.

When it comes to quantum circuit simulations, such representations can only simulate circuits with linear nearest neighbour (LNN) topology, meaning gates have to be applied on neighbouring qubits only. Bond dimensions grow only by application of operators acting on multiple qubits.

\section{Canonicalization}\label{sec:canonicalization}

Casting quantum state vectors or density matrices into their matrix product representations allow for various simplifications. Most of these simplifications use the concept of canonicalization\cite{schollwoeck_density-matrix_2011, orus_practical_2014}, which is a consequence of the SVD or QR operations required for constructing and truncating quantum states in matrix product representations.
Take the SVD of a matrix $M$, we have $M = U \Lambda V^{\dagger}$, where $\Lambda$ is diagonal, $U^{\dagger}U=I$ and $(V^{\dagger}V)^\dagger = VV^\dagger=I$. As such, $U$ is left-normalized, while $V^\dagger$  is right-normalized. Similarly, a QR operation on a matrix $M$ gives $M=QR^\dagger$ with $Q$ and $R^\dagger$ respectively left and right-normalized matrices. By controlling which tensors are left- or right-normalized, many operations can be done on a small subset of tensors, rather than on the entire quantum state.

To illustrate this, let us compute the expectation value of an observable $\hat{O}$ whose support is on $\ket{\sigma_k}$ and $\ket{\sigma_{k+1}}$ for a quantum state $\ket{\psi}$ in the MPS form defined in Eq.~(\ref{eq:mps_repr}):

\begin{align}
	\langle\hat{O}\rangle_{\psi} =&
	\sum_{\substack{\sigma, \sigma', \chi}} A^{[0]\sigma'_0}_{1,\chi_{1}}...A^{[k]\sigma'_k}_{\chi_{k},\chi_{k+1}}A^{[k+1]\sigma'_{k+1}}_{\chi_{k+1},\chi_{k+2}}
	...\nonumber\\&A^{[N]\sigma'_N}_{\chi_{N-1},1}O^{\sigma'_k, \sigma'_{k+1}}_{\sigma_k, \sigma_{k+1}}A^{[0]\sigma_0}_{1,\chi_{1}}...\nonumber\\&
	A^{[k]\sigma_k}_{\chi_{k},\chi_{k+1}}A^{[k+1]\sigma_{k+1}}_{\chi_{k+1},\chi_{k+2}}...A^{[N]\sigma_N}_{\chi_{N-1},1} \langle\boldmath{\sigma'}|\boldmath{\sigma}\rangle
\end{align}

For now the MPS $\ket{\psi}$ is in arbitrary form. We apply QR to each tensor $A^{[i]}$ for $i \in\ [1, k[$ such that $\sum A^{\dagger[i]} A^{[i]} = I$, and similarly apply a RQ for $i \in\ [N, k+1[$ such that $\sum A^{[i]} A^{\dagger[i]} = I$. The state is now in ``canonical form". The left-normalized tensors multiplied with themselves simplify to identity, and similarly for the right-normalized tensors. 

$\hat{O}$ expectation value can then be computed by only multiplying:
\begin{align}
	\allowbreak
	\langle\hat{O}\rangle_{\psi} =
	\sum_{\substack{\sigma, \sigma', \chi}} &A^{[k]\sigma'_k}_{\chi_{k},\chi_{k+1}}A^{[k+1]\sigma'_{k+1}}_{\chi_{k+1},\chi_{k+2}}\nonumber\\&O^{\sigma'_k, \sigma'_{k+1}}_{\sigma_k, \sigma_{k+1}}A^{[k]\sigma_k}_{\chi_{k},\chi_{k+1}}A^{[k+1]\sigma_{k+1}}_{\chi_{k+1},\chi_{k+2}}
\end{align}
The canonical form is needed for efficiently computing the truncation fidelities, and is enforced throughout the entanglement-aware simulation.

\section{Quantum state fidelity}\label{sec:quantum-state-fidelity}

The quantum state fidelity is a measure of the ``closeness" of two quantum states.

For pure states, the fidelity $\mathcal{F}$ between two states $\ket{\psi}$ and $\ket{\phi}$ is defined by the square modulus of their overlap:
\begin{align}
	\mathcal{F}(\psi, \phi) = |\langle\psi|\phi\rangle|^2\label{pure_state_fidelity}
\end{align}

Similarly, Josza\cite{jozsa_fidelity_1994} defined the quantum state fidelity between two mixed states $\rho$ and $\sigma$ as:
\begin{align}
	\mathcal{F}(\rho, \sigma) = Tr\left(\sqrt{\sqrt{\rho}\sigma\sqrt{\rho}}\right)^2\label{josza_fidelity}
\end{align}

However, truncations of mixed states in MPO representation can lead to the loss of their positive semi-definite (PSD) property, which is a necessary condition for computing the matrix square roots in Eq.~(\ref{josza_fidelity}).

An alternative formulation obeying all four of Josza's  axioms\cite{jozsa_fidelity_1994} has been given by X. Wang et al. \cite{wang_alternative_2008} for mixed states. It is defined by:
\begin{align}
	\mathcal{F}(\rho, \sigma) = \frac{|Tr(\rho\sigma)|}{\sqrt{Tr(\rho^2)Tr(\sigma^2)}} \label{alternate_fidelity}
\end{align}

This definition removes the need for the PSD property of the quantum state, and is usable for non-normalized density matrices.

For the rest of the article, the quantum state fidelity for pure state will refer to Eq.~(\ref{pure_state_fidelity}), while it will refer to Eq.~(\ref{alternate_fidelity}) for mixed states.
\section{Truncation fidelity}\label{sec:truncation-fidelity-of-mps}

We call truncation fidelity the quantum state fidelity before and after truncation. We show truncation fidelities can be efficiently computed using only the singular values associated with the bond dimension that is to be truncated.

Let us assume we want to truncate a pure state $\ket{\psi}$ at site $T$, to obtain the truncated state $|\tilde{\psi}\rangle$. The state $\ket{\psi}$ is in canonical form. Contracting both tensors neighbouring the bond dimension $T$, and performing its SVD, we obtain the diagonal matrix $\Lambda$, containing the singular values in decreasing order. Truncating this matrix reduces the overall bond dimension at site $T$, and approximates the state.

Since the state is in canonical form, the truncation fidelity of such truncation is given by:
\begin{align}
	f_T(\tilde{\Lambda}) =	|\langle\psi | \tilde{\psi}\rangle|^2 = \sum_i \Lambda_{ii}^2 \tilde{\Lambda}_{ii}^2 \label{eq:fidelity_pure_state}
\end{align}
with $\Lambda$ and $\tilde{\Lambda}$ the diagonal matrices representing the singular values obtained at site $T$ \emph{before} and \emph{after} truncation respectively.

Similarly, truncating a mixed state $\rho$ in canonical form at site $T$, we obtain the truncated mixed state $\tilde{\rho}$. Using Eq.~(\ref{alternate_fidelity}), and simplifying the tensors by using the canonical simplifications previously described, we obtain:

\begin{align}
	f_T(\tilde{\Lambda}) = \frac{\sum_i \Lambda_{ii}^2\tilde{\Lambda}_{ii}^2}{\sum_i \Lambda_{ii}^2 \sum_i \tilde{\Lambda}_{ii}^2} \label{eq:fidelity_mixed_state}
\end{align}
In both cases, the truncation fidelity $f_T$ depends only on the actual truncated singular values in $\tilde{\Lambda}$. Computing the truncation fidelity is now simply a scalar operation between singular values \emph{before} and \emph{after} truncation.

\section{Entanglement-aware simulations}\label{sec:entanglement-aware-matrix-product-based-simulation-algorithm}

Throughout a matrix product-based simulation algorithm, the quantum state endures successive truncations. The more entangled the state, the bigger the truncation errors, and thus the higher the bond dimensions need to be. As shown in Zhou et. al. article\cite{zhou_what_2020} and further confirmed in \cite{ayral_density-matrix_2022}, the overall quantum fidelity $\mathcal{F}$ between an exact quantum state $\ket{\psi}$ and its truncated counterpart $|\tilde{\psi}\rangle$ can be approximated for noiseless simulations by:
\begin{align}
	\mathcal{F}(n) \approx \prod_{i=1}^{n}f_i \label{eq:product_fidelities}
\end{align}
with $n$ the number of truncations and $f_i$ the individual truncation fidelities obtained at each truncation. This approximation proves remarkably robust at all regimes studied\cite{zhou_what_2020, ayral_density-matrix_2022}.
In the case of noisy systems, Eq.~(\ref{eq:product_fidelities}) underestimates $\mathcal{F}$ since the additional noise reduces the truncation errors committed. In the general case we have:
\begin{align}
	\mathcal{F}(n) \gtrsim \prod_{i=1}^{n}f_i \label{eq:lower_bound_fidelities}
\end{align}


%
%
By enforcing the canonical form at all times throughout the simulation, we can efficiently compute truncation fidelities, and thus have direct access to how close the resulting approximated quantum state is to its exact non-truncated counterpart.
Using Eq.~(\ref{eq:fidelity_pure_state}) for pure states or Eq.~(\ref{eq:fidelity_mixed_state}) for mixed states, we know how many singular values have to be truncated in order to reach a specific target truncation fidelity. Since we know in advance how many 2-qubit gates are to be applied onto the quantum state, we can revert Eq.~(\ref{eq:product_fidelities}) and define the target truncation fidelities $\{f^{(i)}_t\}$ to reach at every truncation so that a fidelity $\mathcal{F}_{min}$ is at least obtained. The bond dimension truncation scheme becomes adaptive.

The simplest way to define the target truncation fidelities $\{f^{(i)}_t\}$ is to define them uniformly based on a fidelity $\mathcal{F}_{min}$ the quantum state has to at least reach:
\begin{align}
	f_t^{(i)}(n) = \mathcal{F}_{min}^{1/n} \quad \forall i \ \in \ [0, n]
\end{align}

\begin{figure}[h]
	\centering
	\includegraphics[width=0.40\textwidth]{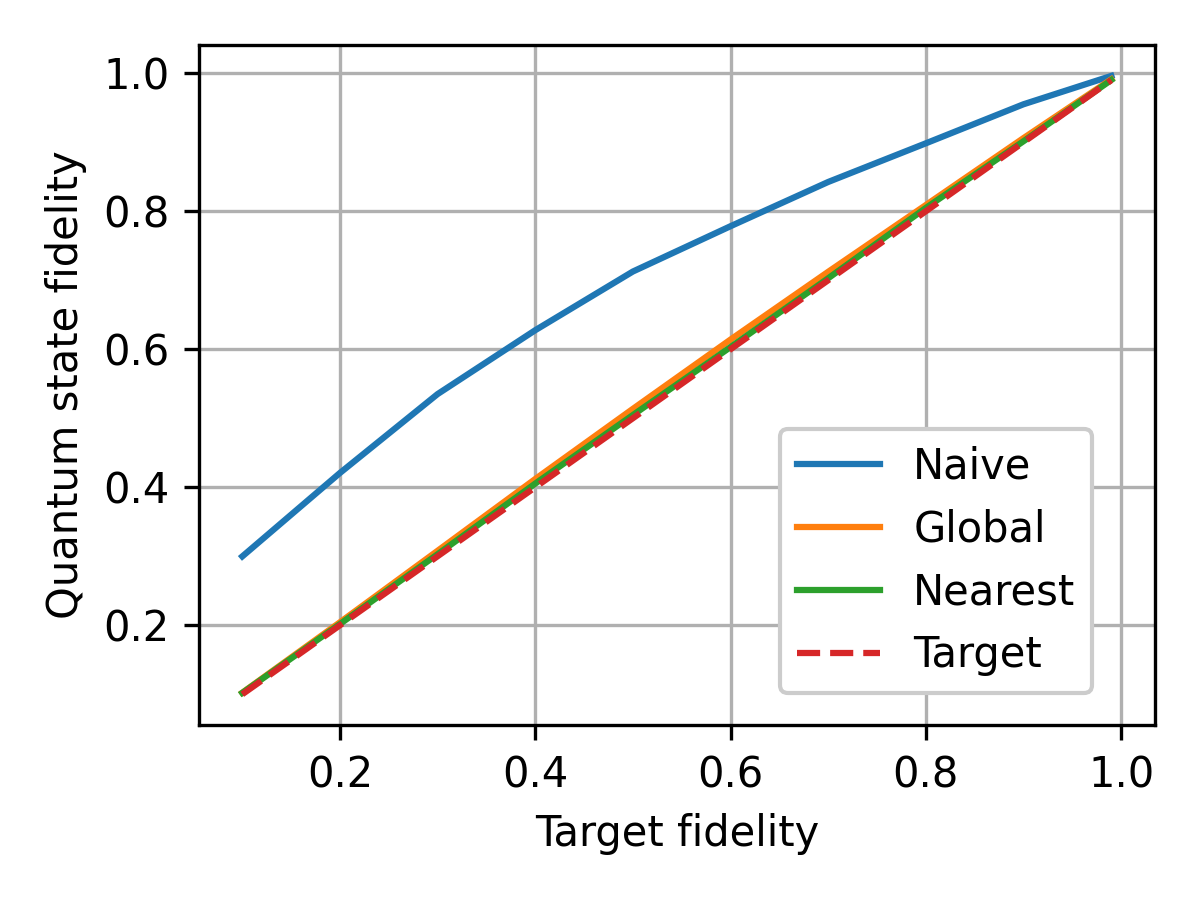} 
	\caption{Comparison between the target fidelity and the actual quantum state fidelity obtained for a 25 qubits Haar-random circuit of depth 20. The ``naive" strategy corresponds to no target truncation fidelity update throughout the simulation. The ``nearest" strategy corresponds to the update of only the upcoming target truncation fidelity, while the ``global" strategy updates all upcoming target truncation fidelities.}
	\label{fig:adaptive_strategy}
\end{figure}

Here, $n$ is the number of 2-qubit gates of the quantum circuit to simulate, i.e. the number of possible truncations.
Throughout the simulation, at every possible truncation, we truncate bond dimensions such that Eq.~(\ref{eq:fidelity_pure_state}) for pure states and Eq.~(\ref{eq:fidelity_mixed_state}) for mixed states are as close to but not lower than $f_t$. In practice the actual truncation fidelity will always be higher than the target truncation fidelity $f_t$, resulting in a final quantum state fidelity higher than the chosen input fidelity $\mathcal{F}_{min}$. To achieve a quantum state fidelity closer to the initial fidelity desired, one can update  target truncation fidelities dynamically after every truncation so that the product of truncation fidelities is as close to $\mathcal{F}_{min}$ as possible, see Fig.~\ref{fig:adaptive_strategy}.


Since bond dimensions now depend on a direct truncation fidelity measure, they increase or decrease based on how they affect the final quantum state fidelity. The smaller the $\mathcal{F}_{min}$ value, the lower the target truncation fidelities, and the more aggressive the truncations will be. As such, the algorithm is ``entanglement-aware" since it is able to detect local entanglement changes, and adapt bond dimensions accordingly.

\begin{figure}[b]
	\centering
	\subfigure[]{\includegraphics[width=0.235\textwidth]{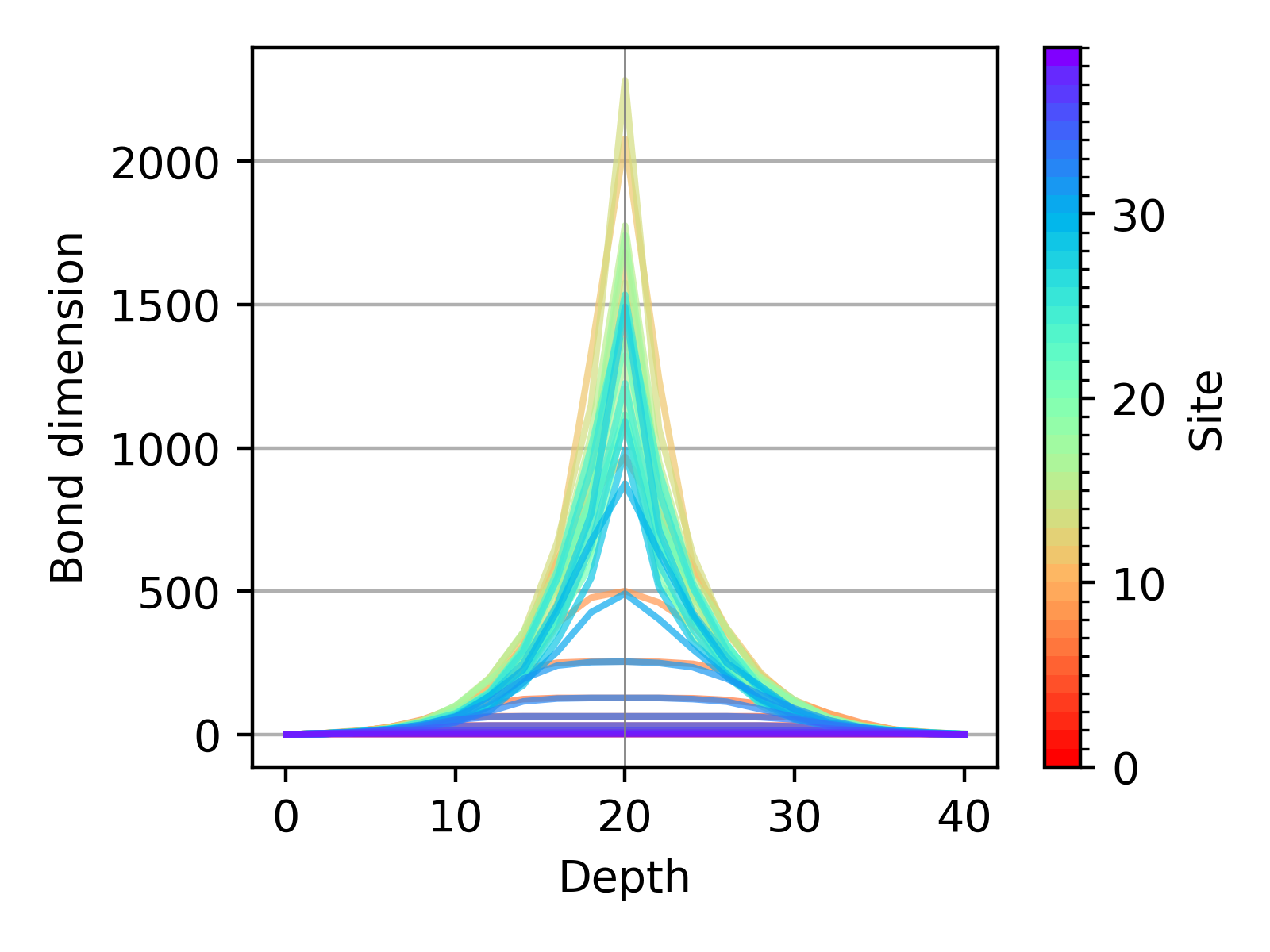}\label{fig:A_entanglement-aware-demonstration}}
	\subfigure[]{\includegraphics[width=0.23\textwidth]{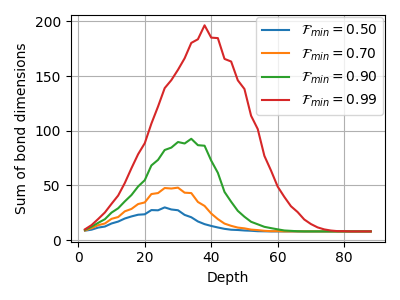}\label{fig:B_entanglement-aware-demonstration}}
	\caption{(\emph{a}) Evolution of the bond dimensions at each MPS site throughout the simulation of a 40 qubits circuit composed of a Haar-random circuit of 20 layers followed by its adjoint. All quantum gates are defined by Haar-random unitary matrices. The bond dimensions increase as the entanglement increases. When the adjoint circuit is reached, entanglement entropy decreases, and the bond dimensions follow. (\emph{b}) Simulation of a noisy quantum circuit of 8 qubits with depolarizing noise parameters $\epsilon_1=\epsilon_2=0.05$ for both one- and two-qubit gates. Noise increases as the depth of the circuit increases. This leads to reduced entanglement needs, which is then exploited by the entanglement-aware algorithm to reduce the state bond dimensions. The lower the desired fidelity $\mathcal{F}_{min}$, the more aggressive the algorithm will be at truncating bond dimensions. Inversely, the higher $\mathcal{F}_{min}$, the more conservative the truncations will be.}
	
\end{figure}

Fig.~\ref{fig:A_entanglement-aware-demonstration} demonstrates this adaptivity on a circuit composed of two subcircuits, the second subcircuit being the adjoint of the first. The initial state is a product state requiring only bond dimensions of 1. In the first subcircuit, as entanglement increases bond dimensions increase, to then decrease in the second subcircuit. At the end of the simulation, the state goes back to the initial state with a bond dimensions of 1 everywhere on the tensor.

In Fig.~\ref{fig:B_entanglement-aware-demonstration}, as entanglement dominates, bond dimensions increase. But eventually noise takes over and reduces the overall state entanglement, leading to decreased bond dimension needs. The lower the desired fidelity, the faster the algorithm is at picking up these changes.

\begin{figure}[h]
	\centering
	\subfigure[]{\includegraphics[width=0.235\textwidth]{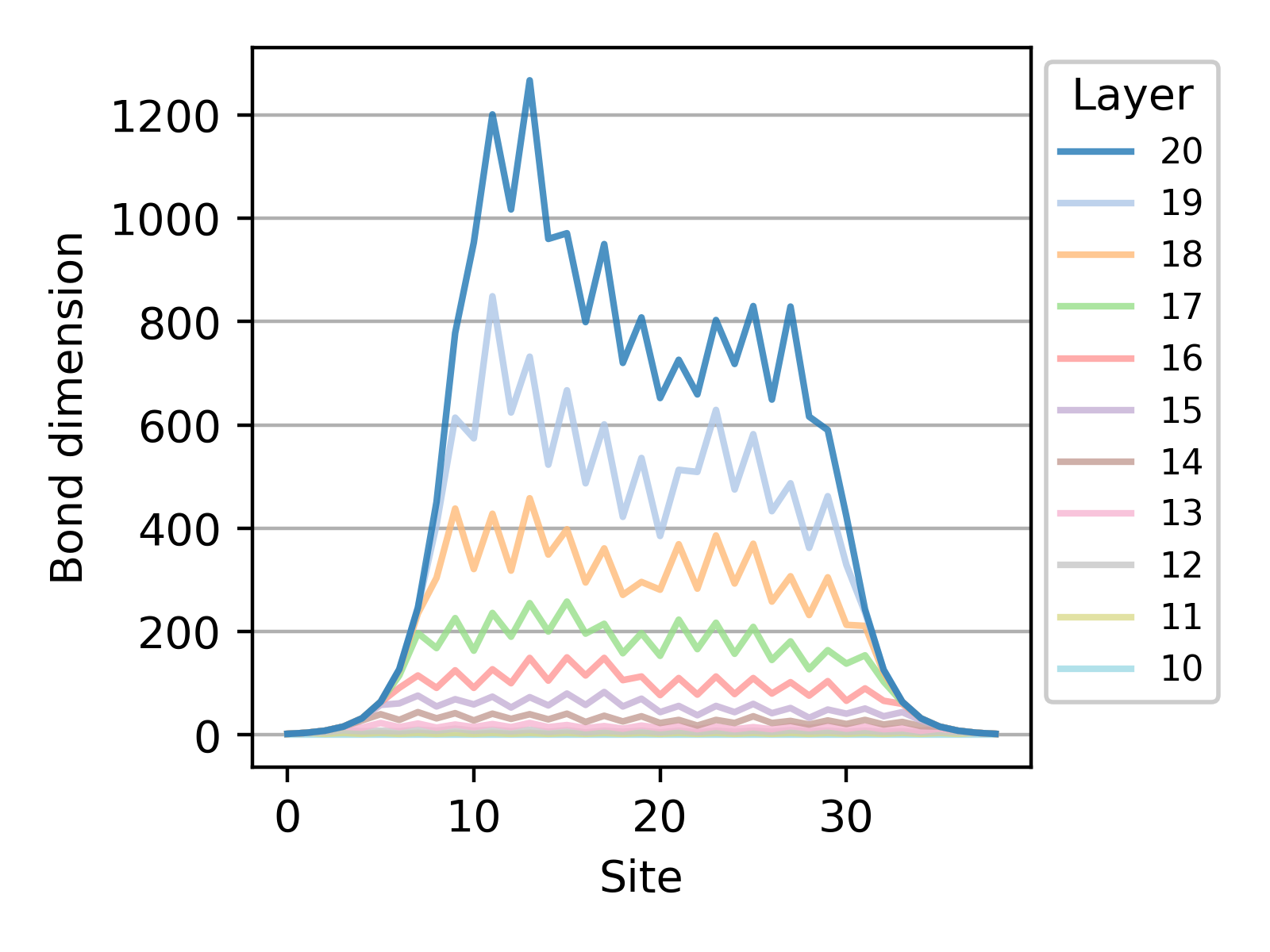}\label{fig:A_intro_figure_fidelity}} 
	\subfigure[]{\includegraphics[width=0.23\textwidth]{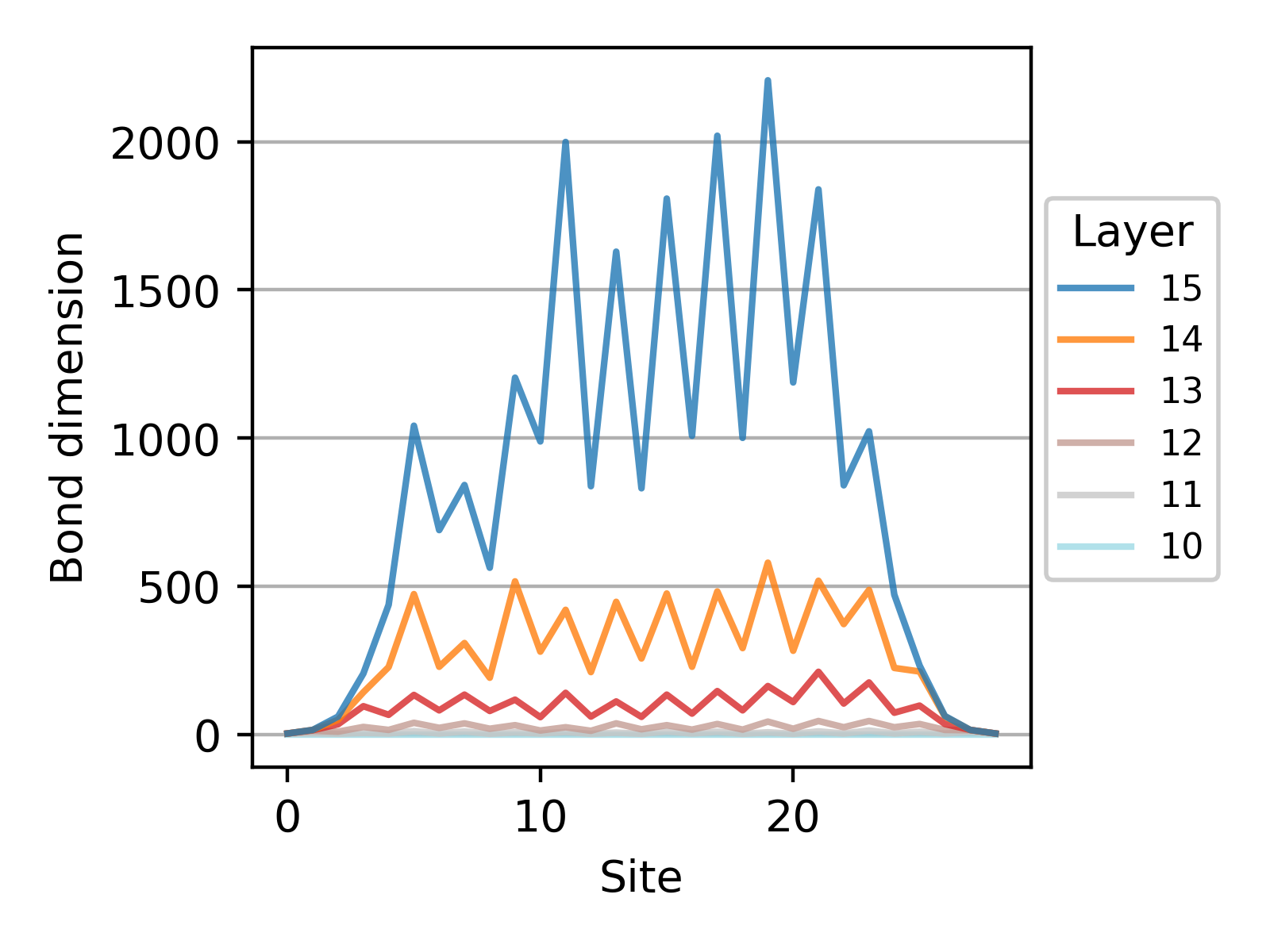}\label{fig:B_intro_figure_fidelity}} 
	
	\caption{(\emph{a}) Simulation of a 40 qubits Haar-random quantum circuit of depth 20 with a target fidelity of $\mathcal{F}_{min}=0.9$. Contrary to intuition, bond dimensions are not higher at the center of the MPS. (\emph{b}) Noisy simulation of a 30 qubits Haar-random quantum circuit of depth 10 with a target fidelity of $\mathcal{F}_{min}=0.9$. Large local entanglement variations are exploited and induce large bond dimensions variations along the MPO.}
	\label{fig:intro_figure_fidelity}
	
\end{figure}

This brings new insights on the dynamics of quantum systems. One could for example expect that the bond dimensions requirements in a Haar-random circuit would be maximal around the middle junction of the MPS. But this is not always the case, as shown in Fig. \ref{fig:A_intro_figure_fidelity}. Noisy systems tend to display large bond dimensions variations that cannot be exploited using fixed maximum bond dimension truncation schemes, see Fig.~\ref{fig:B_intro_figure_fidelity}. This makes adaptive truncation schemes particularly effective for noisy systems.

\begin{figure}[h]
	\centering
	\subfigure[]{\includegraphics[width=0.23\textwidth]{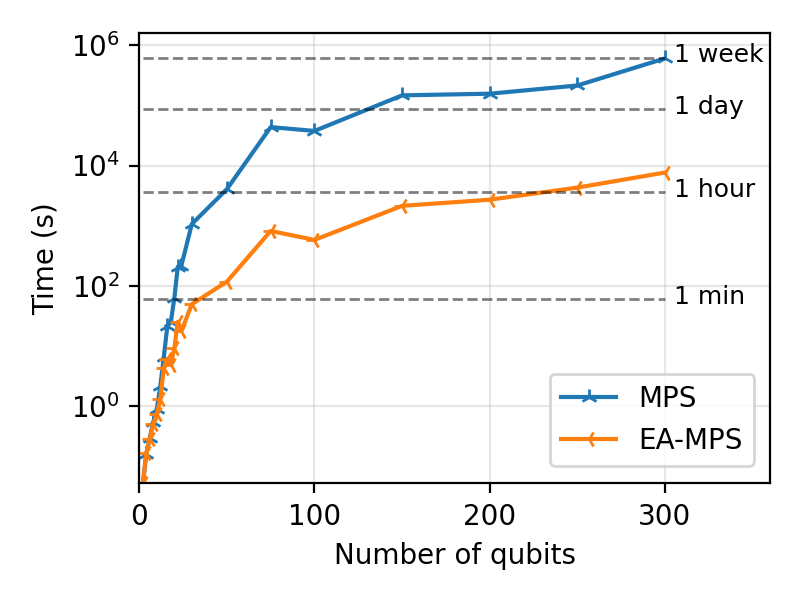}\label{fig:A_benchmarks}} 
	\subfigure[]{\includegraphics[width=0.23\textwidth]{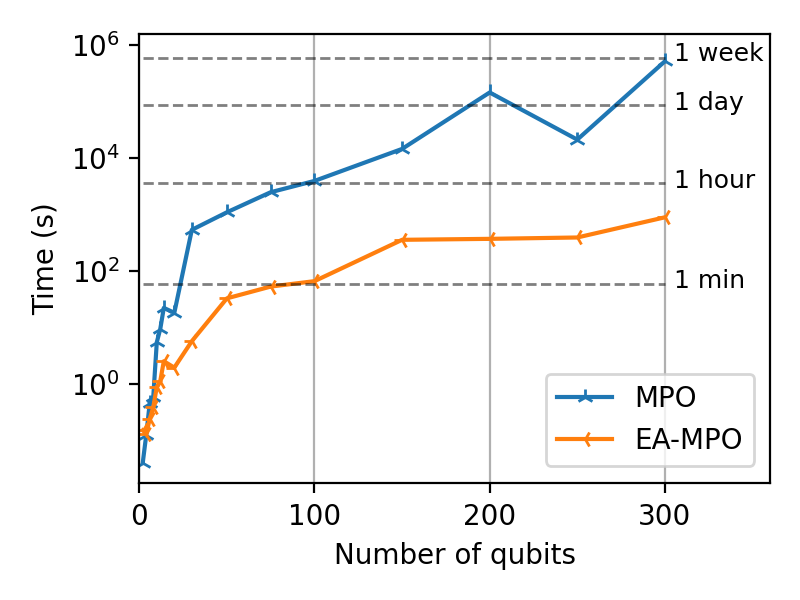}\label{fig:C_benchmarks}}
	\subfigure[]{\includegraphics[width=0.23\textwidth]{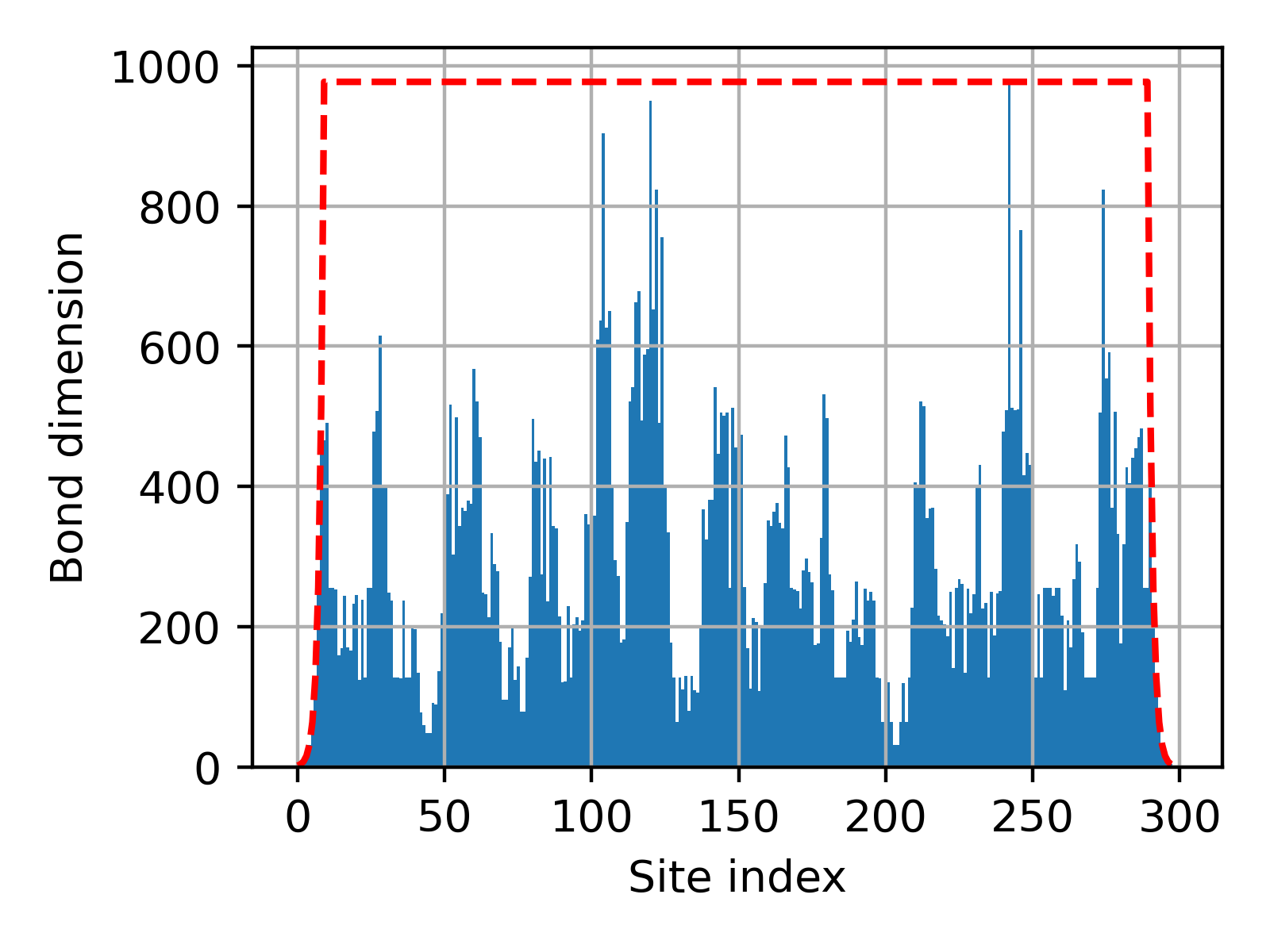}\label{fig:B_benchmarks}}
	\subfigure[]{\includegraphics[width=0.23\textwidth]{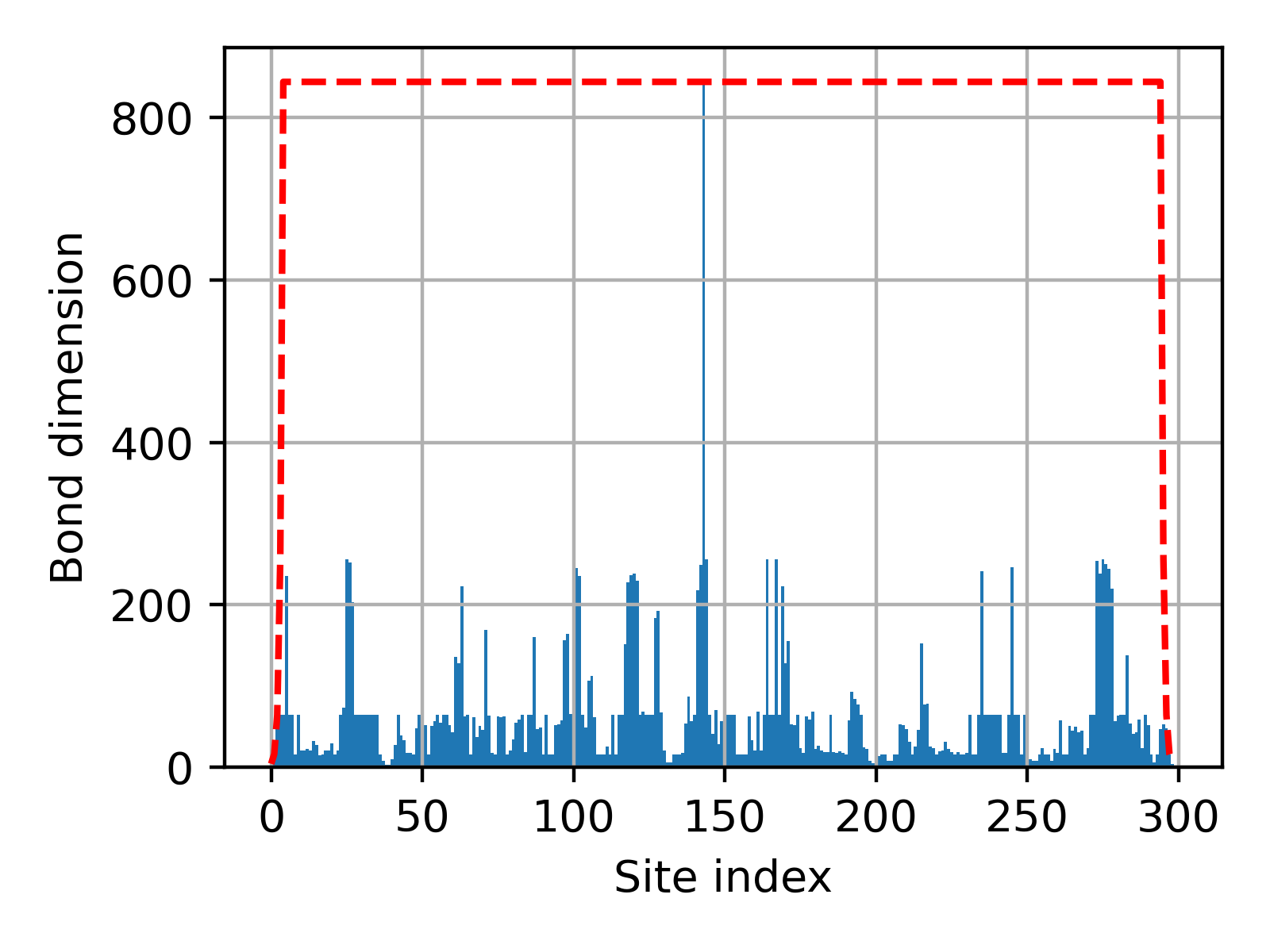}\label{fig:D_benchmarks}}
	\caption{
		(\emph{a}) Simulation time of MPS and EA-MPS for a noiseless random circuit of depth 75. (\emph{b}) Simulation time of MPO and EA-MPO for a noisy random circuit of depth 25. For both figure and for each point, the MPS (resp. MPO) maximum bond dimension is taken as the maximum bond dimension encountered during the EA-MPS (resp. EA-MPO) simulation. This ensures similar final quantum state fidelity $\mathcal{F}_{min}$. For example, for 300 qubits we obtain $\mathcal{F}_{min} = 0.999$ for both regular MPS and MPO, while we obtain $\mathcal{F}_{min} = 0.998$ for EA-MPS and EA-MPO.
		(\emph{c}) and (\emph{d}) show the final state bond dimensions for the 300 qubits data point in (\emph{a}) and (\emph{b}) respectively. The blue lines show the bond dimensions of the state obtained through EA simulation, while the red dotted lines are the bond dimensions obtained through regular fixed bond dimension simulation.
	}
	
\end{figure}

Benchmarks of noiseless and noisy entanglement-aware simulations are presented in Fig.~\ref{fig:A_benchmarks} and Fig.~\ref{fig:C_benchmarks}, and compared to standard methods. The random circuit used is defined based on Cheng's article\cite{cheng_simulating_2021}, which alternates randomly chosen layers of one- and two-qubit gates. The standard MPS simulation requires one week of simulation time for a 300 qubits random circuit of depth 75, while the entanglement-aware version only needs 2 hours. For a depth of 25, the standard MPO simulation requires also one week, but EA-MPO only needs 40 minutes.
Fig.~\ref{fig:B_benchmarks} and Fig.~\ref{fig:D_benchmarks} show the bond dimensions of the resulting approximate quantum state for 300 qubits. 
One has to keep in mind that, assuming it is possible to guess which maximum bond dimension was needed in the first place, standard MPS and MPO simulation algorithms have to simulate the state using very large bond dimensions over the entire tensor network for a final quantum state fidelity gain lower than $0.001$. Note also that while it may be tempting to truncate some of the high peaks displayed in Fig.~\ref{fig:B_benchmarks} or Fig.~\ref{fig:D_benchmarks}, it is exactly these truncations which would lead to the largest fidelity losses on the final quantum state, since it is where the entanglement is highest, and thus where the truncation errors are maximal.

Both EA-MPS and EA-MPO simulation algorithms vastly outperform regular MPS and MPO simulations in all cases studied, may it be in terms of quantum state fidelity, computation time or memory footprint.

It is however difficult to assess the exact scaling improvement over regular fixed maximum bond dimension methods, as it is problem dependent. Also, as bond dimensions are allowed to grow indefinitely, a quantum circuit inducing too much entanglement will remain almost impossible to simulate. For that reason, it is advisable to cap adaptive bond dimensions up to a maximum value which, if reached, enforces bond dimension truncations. In that case, the desired fidelity can no longer be guaranteed.


\section{Conclusion}
In conclusion, I have shown that an adaptive bond dimension method far outperforms methods based on arbitrarily chosen maximum bond dimensions. The metric used in this article is the quantum state fidelity because of its straightforward relationship with truncation fidelities, but other metrics could be used.
This method only adds the computation of the truncation fidelities, but this additional cost is largely compensated for by the overall bond dimension reduction it allows. Since bond dimensions growth can still be constrained by a maximum bond dimension, and because the loss in final quantum state fidelity is negligible, I expect adaptive truncation methods to push even further the quantum systems simulation capabilities of tensor network-based methods on classical hardware.

\bibliography{sn-bibliography}

\end{document}